\documentclass[aps,prl,twocolumn,superscriptaddress,showpacs,preprintnumbers]{revtex4-1}
\usepackage[colorlinks=true, pdfstartview=FitV, linkcolor=red, citecolor=blue, urlcolor=black, pdftitle={},pdfauthor={Tomoya Hayata},pdfsubject={}, pdfkeywords={}]{hyperref}
\usepackage[pdftex]{graphicx}
\usepackage{setspace,bm,times}
\usepackage{latexsym,amssymb,amsmath,mathrsfs}
\usepackage{color}
\graphicspath{{./figure/}}

\newcommand{\cA}{{\cal A}}

\newcommand{\hi}{\hat{i}}
\newcommand{\hx}{\hat{x}}
\newcommand{\hy}{\hat{y}}
\newcommand{\hz}{\hat{z}}
\newcommand{\vphi}{\varphi}

\newcommand{\be}{\begin{equation}}      
\newcommand{\ee}{\end{equation}}      
\newcommand{\bea}{\begin{eqnarray}}      
\newcommand{\eea}{\end{eqnarray}}

\begin{document}
\title{Chiral magnetic effect by synthetic gauge fields}
\author{Tomoya Hayata}
\affiliation{Department of Physics, Chuo University, Tokyo 112-8551, Japan}
\author{Masahito Ueda}
\affiliation{Department of Physics, The University of Tokyo, Tokyo 113-0031, Japan}
\affiliation{RIKEN Center for Emergent Matter Science (CEMS), Wako, Saitama 351-0198, Japan}

\date{\today}

\begin{abstract}
We study the dynamical generation of the chiral chemical potential in a Weyl metal constructed from a three-dimensional optical lattice and subject to synthetic gauge fields. 
By numerically solving the Boltzmann equation with the Berry curvature in the presence of parallel synthetic electric and magnetic fields, 
we find that the spectral flow and the ensuing chiral magnetic current emerge.
We show that the spectral flow and the chiral chemical potential can be probed by time-of-flight imaging.

\end{abstract}

\pacs{67.85.-d,03.65.Vf,11.30.Rd,47.11.-j}
\maketitle

{\it Introduction.} 
Topological states of matter have attracted growing attention in recent years.
The Berry phase and curvature~\cite{Berry45} provide a universal understanding of anomalous transports in such states~\cite{RevModPhys.82.1539,RevModPhys.82.1959}.
The prime example is the quantum Hall effect in two-dimensional electron systems~\cite{PhysRevLett.45.494,PhysRevB.23.5632}, 
where the quantized Hall conductance is characterized by the first Chern numbers which are expressed in terms of the Berry phase~\cite{PhysRevLett.49.405,PhysRevLett.51.51,PhysRevB.31.3372}. 
Recently, the Berry phase has been applied to study a nondissipative current in chiral systems~\cite{Stephanov:2012ki,Son:2012wh,Son:2012zy,Chen:2012ca}.
It has been shown that the Berry curvature on the Fermi surface of Weyl fermions has a close connection with the triangle anomaly and leads to a nondissipative current induced by external magnetic fields.
Such an effect was originally proposed in Refs.~\cite{Vilenkin:1980fu,Nielsen:1983rb}, 
and later it has been applied to explain charge-dependent azimuthal correlations in relativistic heavy-ion collision experiments and  termed the chiral magnetic effect~\cite{Kharzeev:2007jp,Fukushima:2008xe,Abelev:2009ac,Abelev:2009ad}.

The chiral magnetic effect has been actively investigated also in condensed-matter materials under the name of Weyl semimetals~\cite{Xu613,Lu622,PhysRevX.5.031013}, which are three-dimensional analogues of graphene.
In such a system, Weyl fermions (nodes) are realized as band touching points with a definite topological character~\cite{Murakami,PhysRevB.83.205101,PhysRevLett.107.127205}; the effective Hamiltonian near a Weyl node becomes that of a Weyl fermion in relativistic theory. 
The Weyl nodes act as monopoles in momentum space, 
and naturally exhibit topological properties described by the Berry curvature.
The key signal of the chiral magnetic effect, i.e., a negative and anisotropic magnetoresistance~\cite{PhysRevB.88.104412} has been experimentally observed~\cite{PhysRevX.5.031023}.

Contrary to the quantum Hall effect, the chiral magnetic effect arises only in nonequilibrium. 
It requires the difference between the Fermi surfaces of right- and left-handed Weyl fermions, which cannot be realized in equilibrium~\cite{PhysRevB.89.035142}. 
The chiral chemical potential, which quantifies the difference, is only dynamically generated.
The mechanism for the dynamical generation of the chiral chemical potential awaits full understanding, which is crucially important for the study of anomaly induced transport.

Ultracold atom gases are ideally suited to investigate such nonequilibrium physics of interacting particles~\cite{RevModPhys.80.885,RevModPhys.83.863}. 
For example, the long-time dynamics towards thermalization has been experimentally observed in one-dimensional systems~\cite{2012NatPh...8..325T}.
Even though atoms are neutral and do not interact with electromagnetic fields, we can simulate anomalous transport induced by them by using ultracold atoms thanks to the invention of synthetic gauge fields~\cite{Dalibard:2010ph,2014RPPh...77l6401G}.  
Furthermore, by exploiting a Feshbach resonance~\cite{inoue,PhysRevLett.81.69,RevModPhys.80.885,RevModPhys.82.1225}, we can change the magnitude of the coupling strength to study the physics of quantum anomalies in strongly-correlated systems. 

In this Letter, we study the dynamical generation of the chiral chemical potential in a three-dimensional optical lattice system with Weyl nodes.
We numerically solve the time evolution of the Boltzmann equation with the Berry curvature in the presence of parallel synthetic electric field $\bm{E}$ and magnetic field $\bm{B}$.
We show that the excitation from the left-handed Weyl nodes to the right-handed ones occurs only if $\bm{E}\cdot\bm{B}\neq0$, 
which is referred to as the spectral flow.
We discuss how to experimentally observe the spectral flow.
The dynamics of the chiral magnetic current is also discussed.

\begin{figure}[t]
\centering
\includegraphics[scale=0.4]{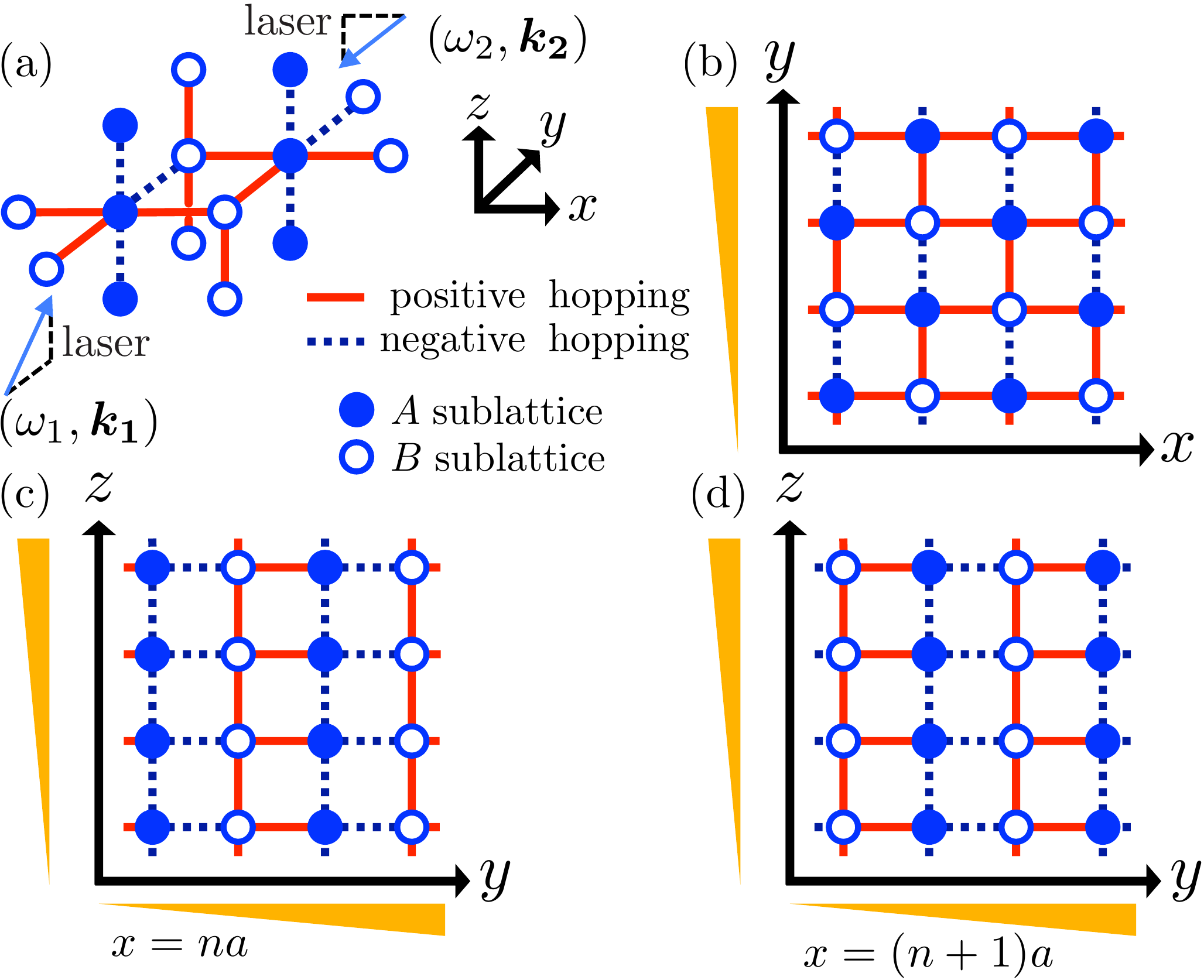}
\caption{\label{fig:lattice}
Schematic illustration of the present optical lattice system.
(a) Three-dimensional sketch. 
A pair of Raman lasers is illustrated by arrows.
(b) Schematic illustration of the Harper Hamiltonian in the $x$-$y$ plane.
The orange triangle denotes the direction along which optical potentials are tilted.
(c) and (d) Alternating two-dimensional lattices in the $y$-$z$ plane.
}
\end{figure}

{\it Model.}
We consider a spinless fermion in a three-dimensional optical lattice.
We adopt the cubic lattice system proposed in Ref.~\cite{PhysRevLett.114.225301}, 
which is constructed  by stacking Harper Hamiltonians~\cite{0370-1298-68-10-304,PhysRevB.14.2239} along the third direction.
The Hamiltonian is 
\be
\begin{split}
 &{\cal H} 
= -\frac{1}{2}\sum_{\bm{A}}\; K_x b^\dagger_{\bm{A}+\hx} a_{\bm{A}} -K_y b^\dagger_{\bm{A}+\hy} a_{\bm{A}}-K_z a^\dagger_{\bm{A}+\hz} a_{\bm{A}} 
\\
& -\frac{1}{2}\sum_{\bm{B}}\; K_x a^\dagger_{\bm{B}+\hx} b_{\bm{B}} +K_y a^\dagger_{\bm{B}+\hy} b_{\bm{B}}+K_z b^\dagger_{\bm{B}+\hz} b_{\bm{B}} 
+{\rm h. c.}, 
\end{split}
\label{eq:Hamiltonian}
\ee
where $a^\dagger_{\bm{A}}$ and $a_{\bm{A}}$ ($b^\dagger_{\bm{B}}$ and $b_{\bm{B}}$) are the creation and annihilation operators of fermions at $A$ ($B$) sites (see Fig.~\ref{fig:lattice}), $K_i$ denotes the hopping parameter along the $i$ direction ($i=x, y, z$), and $\hi$ denotes a unit vector in the $i$ direction.
The site $\bm{r}$ is labeled by integers $(m,n,l)$ as $\bm{r}=(x,y,z)=(m\hx+n\hy+l\hz)a$ with $a$ being the lattice spacing, 
and belongs to the $A$ ($B$) sublattice if $m+n$ is odd (even).
The choice of hopping amplitudes is illustrated in Fig.~\ref{fig:lattice}.
The position-dependent hopping means that the magnetic flux per plaquette $\vphi$ is nonzero ($\vphi=\pi$), which has experimentally been realized by laser-assisted tunneling~\cite{PhysRevLett.107.255301,PhysRevLett.111.185301,PhysRevLett.111.185302} or shaking of an optical lattice~\cite{PhysRevLett.108.225304} for the case of staggered magnetic flux.

The Hamiltonian~\eqref{eq:Hamiltonian} is written in the wave-number basis as ${\cal H} =\int d^3p\psi^\dagger_{\bm{p}} h(\bm{p})\psi_{\bm{p}}$ with
$ h(\bm{p}) =-K_x \cos(p_xa)\sigma_x-K_y\sin(p_ya)\sigma_y+K_z \cos(p_za)\sigma_z$, where 
we call sublattice indices ``spin" indices and introduce the following pseudo-spin representation: $\psi^T_{\bm{p}}=(a_{\bm{p}},b_{\bm{p}})^T$. 
The energy eigenvalue is 
$ E_{\pm} = \pm\sqrt{K_x^2\cos^2(p_xa)+K_y^2\sin^2(p_ya)+K_z^2\cos^2(p_za)}$, 
 which has eight Weyl points, $\bm{p}_w=(\pm\pi/(2a),0,\pm\pi/(2a))$ and $(\pm\pi/(2a),\pi,\pm\pi/(2a))$ in the first Brillouin zone.
We can assign the chirality $\kappa=+1$ or $-1$ to the Weyl points, depending on the sign of  $v_xv_yv_z$ with $v_i=\partial E_+/\partial p_i$ at $\bm{p}=\bm{p}_w$.
The Berry connection is defined by 
$ i \bm{\cA}^{\pm}(\bm{p})=   \left(u^{\pm}(\bm{p})\right)^\dagger\nabla_{\bm{p}}u^{\pm}(\bm{p})$ with $u^{\pm}(\bm{p})$ being the wave function of a positive (negative) energy eigenstate.
We then find a nonzero Berry curvature 
\be
\begin{split}
&\bm{\Omega}^{\pm}(\bm{p}) \equiv\nabla_{\bm{p}}\times \bm{\cA}^{\pm} \\
&=\pm a^2\frac{K_xK_yK_z}{2E_+^3}
\begin{pmatrix}
-\cos(p_xa)\cos(p_ya)\sin(p_za) \\
+\sin(p_xa)\sin(p_ya)\sin(p_za) \\
-\sin(p_xa)\cos(p_ya)\cos(p_za)
\end{pmatrix} .
\end{split}
\label{eq:weyl_monopole}
\ee
A surface integration of $\bm{\Omega}^{\pm}$ becomes
$\int d\bm{S}_p\cdot\Omega^{\pm}(p)=\pm2\pi\kappa$,
where the integration is performed over the surface that encloses only one of the Weyl nodes.
We can interpret $\bm{\cA}^{\pm}(p)$ and $\bm{\Omega}^{\pm}(p)$ as the vector potential and the magnetic field in momentum space, respectively.
The ``magnetic field" is generated by (anti-)monopoles at the Weyl nodes.
We remark that the Berry curvature affects particles occupying the positive and negative energy eigenstates in an opposite manner.

We need to apply further synthetic electric and magnetic fields.
The magnetic field is already embedded in the Hamiltonian~\eqref{eq:Hamiltonian}.
To simulate the Weyl fermion at finite magnetic fields, it is enough to slightly change the momentum of Raman lasers.
On the other hand, to create a synthetic electric field, we have to apply a time-dependent phase simultaneously with the position-dependent phases induced by laser-assisted tunneling.
At finite magnetic fields, the energy is modified because a quasi-particle has a nonzero magnetic moment.
The corrected energy reads $ \varepsilon_{\pm} =\pm E_+\left(1- \frac{e}{\hbar c}\bm{B}\cdot\bm{\Omega}^{+}\right)$~\cite{RevModPhys.82.1959},
which is used in the kinetic equation discussed below.

{\it Chiral kinetic theory.} 
We numerically solve the collisionless Boltzmann equation by assuming the weak-coupling and dilute limit. 
We consider the Wigner function in the pseudospin representation $n(\bm{x},\bm{p})=\int d^3y e^{i\bm p\cdot \bm y}\langle\psi^\dagger({\bm x}+{\bm y}/2)\psi({\bm x}-{\bm y}/2)\rangle$, which serves as a distribution function in the phase space ($\bm{x},\bm{p}$).
According to the Liouville theorem $dn/dt=0$, the collisionless Boltzmann equation reads
\be
\partial_t n+\dot{\bm x}\cdot\nabla_{\bm x} n+\dot{\bm p}\cdot\nabla_{\bm p} n=0 ,
\label{eq:boltzmann}
\ee
with
\bea
\sqrt{\omega}\dot{\bm x} &=& \bm v+\frac{e}{\hbar}\bm E\times\bm \Omega+\left(\bm v\cdot\bm \Omega\right)\frac{e}{\hbar c}\bm B , 
\label{eq:eom1} \\
\sqrt{\omega}\dot{\bm p} &=& \frac{e}{\hbar}\bm E+\bm v\times\frac{e}{\hbar c}\bm B+\left(\frac{e}{\hbar}\bm E\cdot\frac{e}{\hbar c}\bm B\right)\bm \Omega ,
\label{eq:eom2}
\eea
where $\omega=(1+e\bm B\cdot\bm \Omega/(\hbar c))^2$ and $\bm v=\nabla_{\bm p} \varepsilon_+/\hbar$ is the velocity of a quasiparticle~\cite{Stephanov:2012ki,Son:2012wh,Son:2012zy,Chen:2012ca,PhysRevB.88.104412}.
This equation can be derived from quantum field theory on the basis of the derivative expansion of the Wigner function~\cite{Son:2012zy,Chen:2016fns}.

Since we are interested in the momentum distribution function, 
we first integrate Eq.~\eqref{eq:boltzmann} over $\bm{x}$ and solve ($1+3$)-dimensional equation of $n_{\bm p}=\int d^3xn(\bm{x},\bm{p})/V$ with $V$ being the volume in real space:
\be
\partial_t n_{\bm p}+\dot{\bm p}\cdot\nabla_{\bm p} n_{\bm p}=0 .
\label{eq:boltzmann_p}
\ee
We emphasize that the reduction is exact as long as electric and magnetic fields are spatially uniform.
Because of the Berry curvature in Eq.~\eqref{eq:eom2}, the momentum distribution isotropically expands (contracts) according to Eq.~\eqref{eq:boltzmann_p} if $\bm{E}\cdot\bm{B}$ is nonzero. 
Then the Fermi surface of the right-handed (left-handed) Weyl node enlarges (shrinks) and the difference between the Fermi surfaces (i.e, the chiral chemical potential) is dynamically generated.

\begin{figure}[t]
\centering
\includegraphics[scale=0.4]{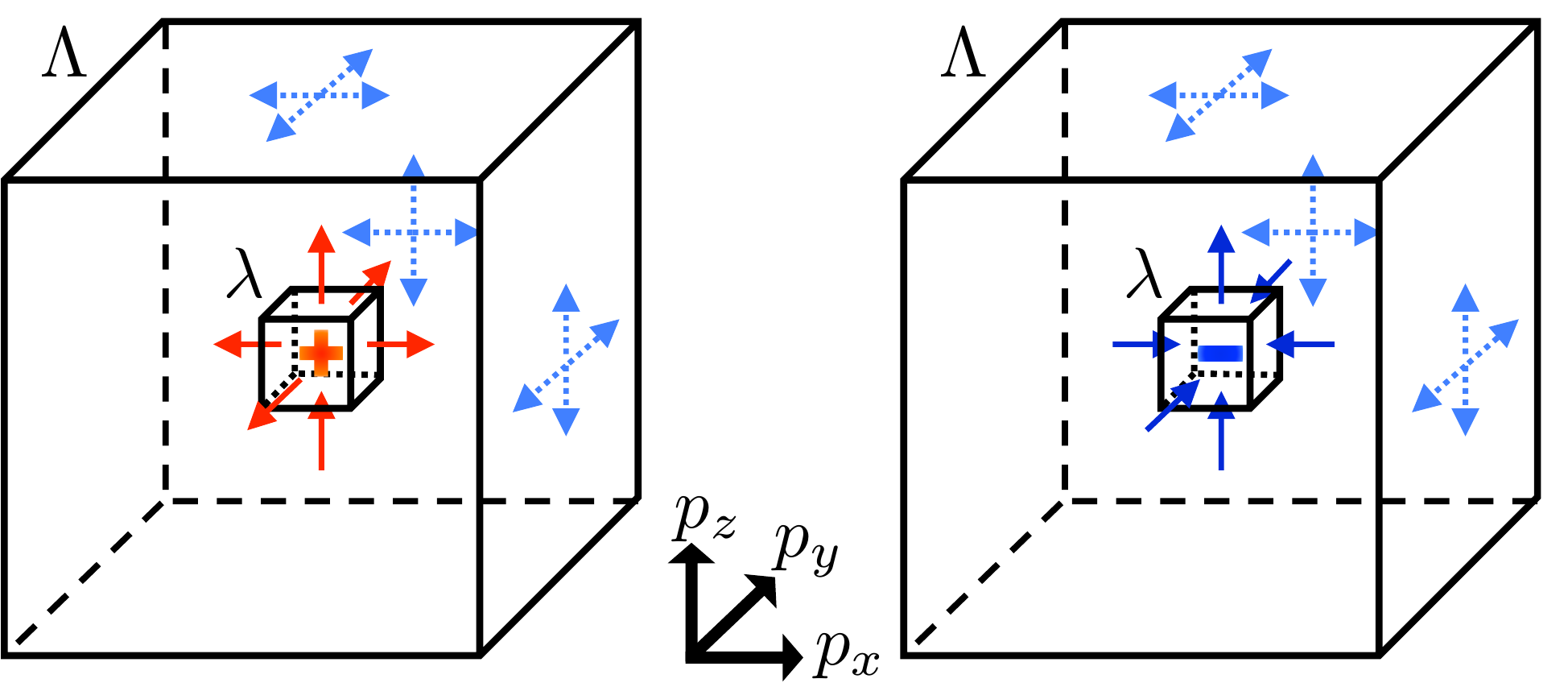}
\caption{\label{fig:boundary}
Boundary conditions of the Boltzmann equation. 
The right- and left-handed Weyl nodes are located at the centers of the cubes. 
The momentum-space fluxes are illustrated by arrows.
The fluxes perpendicular to the outer boundaries are set to zero.
The fluxes that enter or leave the inner boundaries are fixed by the divergence of a monopole or an anti-monopole.
}
\end{figure}

{\it Numerical simulation.} 
We numerically solve Eq.~\eqref{eq:boltzmann_p} by adopting the constrained interpolation profile (CIP) scheme~\cite{Yabe2001556}, 
which is used to solve the Boltzmann equation (the Vlasov-Maxwell or the Vlasov-Poisson equation) stably and accurately in plasma physics and astrophysics.

We consider the synthetic electric and magnetic fields along the $z$ direction, $\bm{E}=(0,0,E)$ and $\bm{B}=(0,0,B)$.
We set $K_x=K_y=K_z=K$, $aeE/K=(\Theta(1-K t/\hbar)-\Theta(-Kt/\hbar))/200$ with $\Theta$ being the Heaviside step function, 
and $\vphi_r\equiv\vphi/\vphi_0=1/5000$ ($\vphi=a^2 B$) in terms of flux per unit cell with the flux quanta $\vphi_0=2\pi \hbar c/e$.
As an initial state of $n_{\bm p}$, we choose the Fermi distribution with temperature $T/K=1/400$ and chemical potential $\mu/K=1/10$.
We perform numerical simulations at a sufficiently low temperature such that the Fermi surface is well defined.
Also we choose a small chemical potential so that the distribution is well localized around each Weyl node.
Then instead of solving the Boltzmann equation over the entire momentum space, 
we have to solve it only near the Weyl node.
We consider a three-dimensional cube defined by $|p_{i=x,y,z}-p_{wi}|<\Lambda$, 
and solve the time evolution of $n_{\bm p}$ only inside of the cube to get better spatial resolution.
We choose $\bm{p}^{\pm}_w=(\pi/(2a),0,\pm\pi/(2a))$ out of the eight Weyl nodes, which have the positive and negative chiralities. 
The distribution around other Weyl nodes can be obtained simply by shifting the data shown below.
The boundary conditions are schematically illustrated in Fig.~\ref{fig:boundary}.
We adopt the slip-free boundary conditions for the outer boundaries. 
We also need to impose boundary conditions at the deep inside of the cube since the Berry curvature diverges at the Weyl node, 
where the kinetic description apparently breaks down. 
Following Ref.~\cite{Stephanov:2012ki}, we fix the distribution with the initial equilibrium value inside of the small cube defined by $|p_i-p_{wi}|<\lambda\ll \Lambda$. 
The momentum-space fluxes that enter or leave the inner boundaries are given by the divergence of a monopole or an anti-monopole.  
We have confirmed that the following results are independent of $\Lambda$ and $\lambda$. 

\begin{figure}[t]
\centering
\includegraphics[scale=0.75]{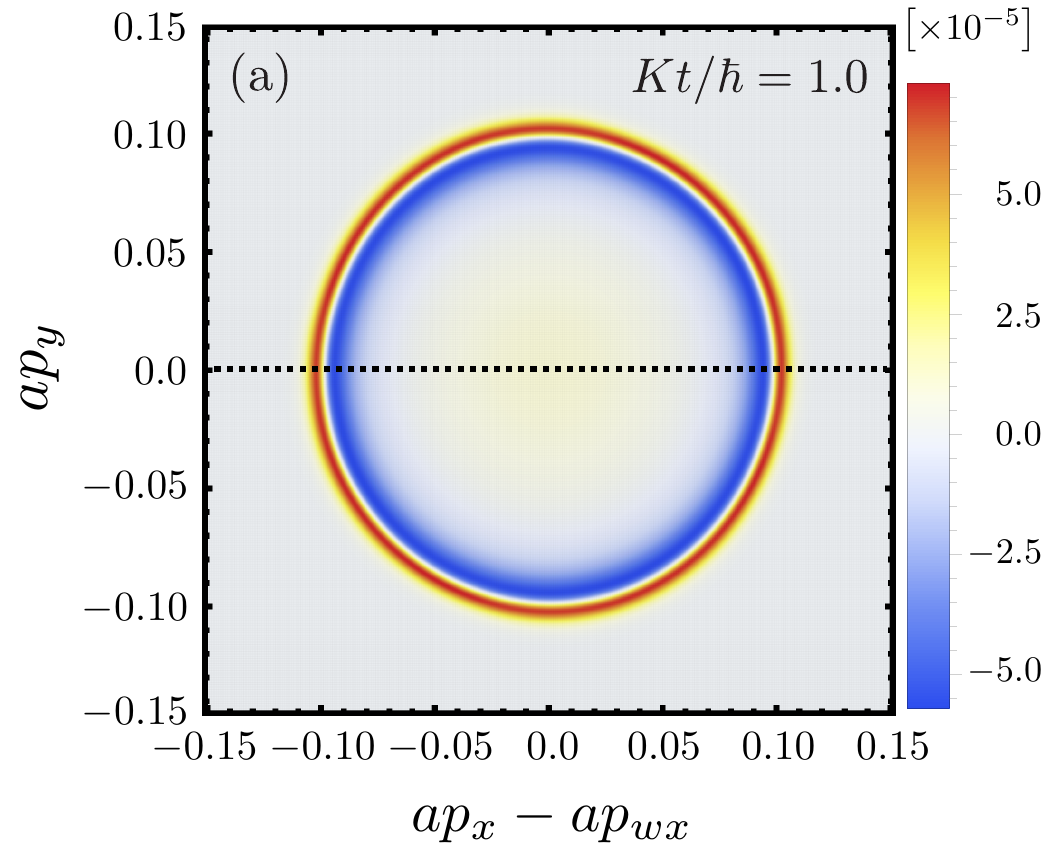}
\centering
\includegraphics[scale=0.6]{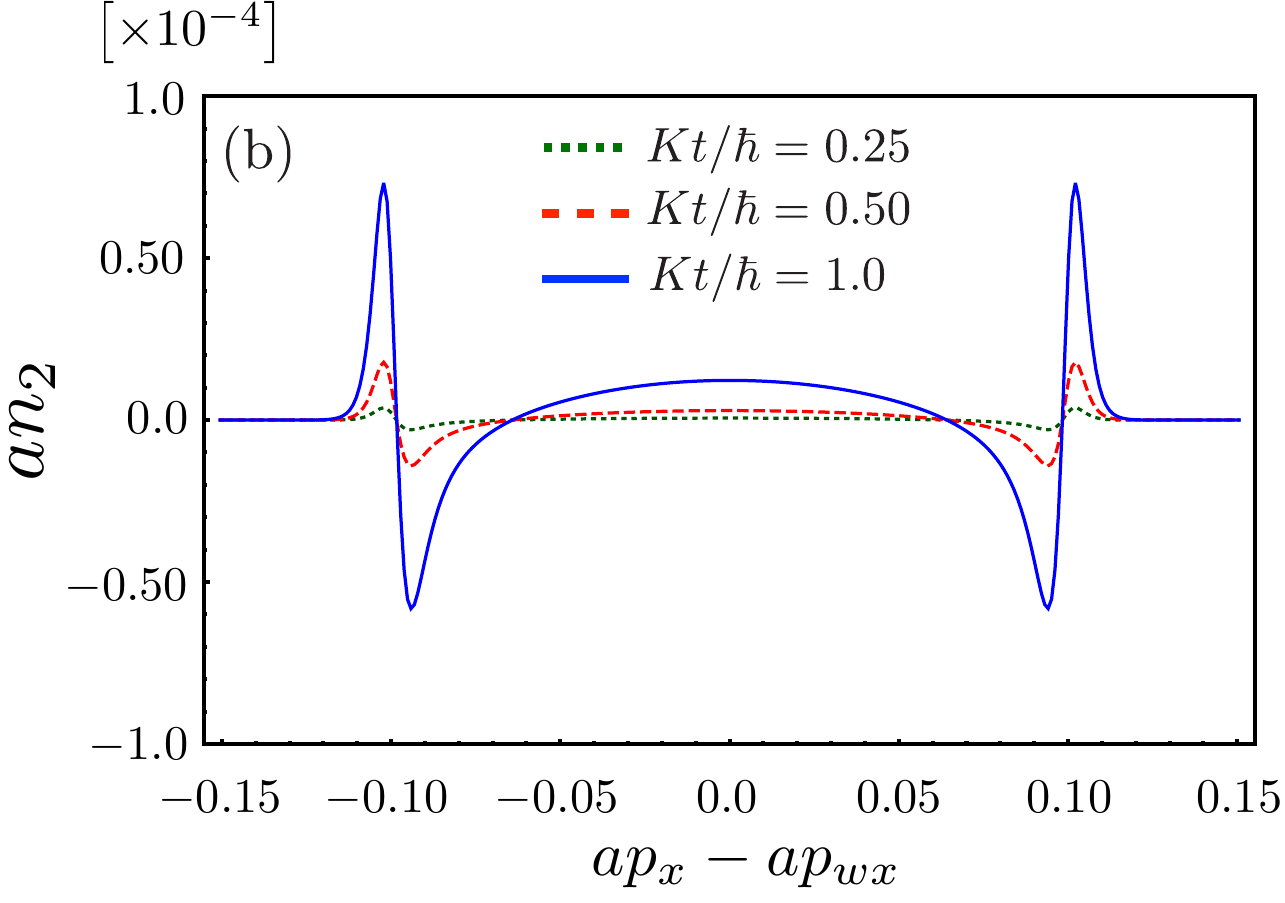}
\caption{\label{fig3:ring}
(a) Momentum distribution $a\delta n_2(p_x,p_y)$ at $Kt/\hbar=1.0$.
(b) One-dimensional distribution $a\delta n_2(p_x,p_{wy})$ along the dashed line in (a) at $Kt/\hbar=0.25$, $0.50$, and $1.0$.
}
\end{figure}

We show the momentum distribution integrated over $p_z$, $n_2(t,p_x,p_y)=\int dp_z \sqrt{\omega}n_{\bm p}$, 
which can experimentally be measured by time-of-flight imaging.
To see the spectral flow clearly,  we show the deviation from the initial equilibrium distribution $\delta n_2(t,p_x,p_y)=n_2(t,p_x,p_y)-n_2(0,p_x,p_y)$ in Fig~\ref{fig3:ring}(a).
We find that positive and negative rings appear just above and below the initial Fermi surface $\varepsilon_F/K\sim a\sqrt{(p_x-p_{wx})^2+p_y^2}\sim \mu/K=1/10$,
and the difference between the Fermi surfaces of the left- and right-handed Weyl nodes is dynamically generated. 
The time dependence of the double-ring pattern is shown in Fig~\ref{fig3:ring}(b).
We expect that this pattern is robust and can be observed through absorption imaging after time-of-flight ballistic expansion with adiabatically ramping down the lattice potential and mapping the lattice momentum to the free-particle one~\cite{PhysRevLett.87.160405,PhysRevLett.94.080403}.

The chiral chemical potential is in general small compared with the chemical potential. Its calculation requires a fine momentum resolution near the Fermi surface
and hence a huge computational cost to keep the required resolution over the entire momentum space. 
Also it is impractical since the distribution is exponentially small in most of the momentum region away from the Fermi surface. 

\begin{figure}[t]
\centering
\includegraphics[scale=0.54]{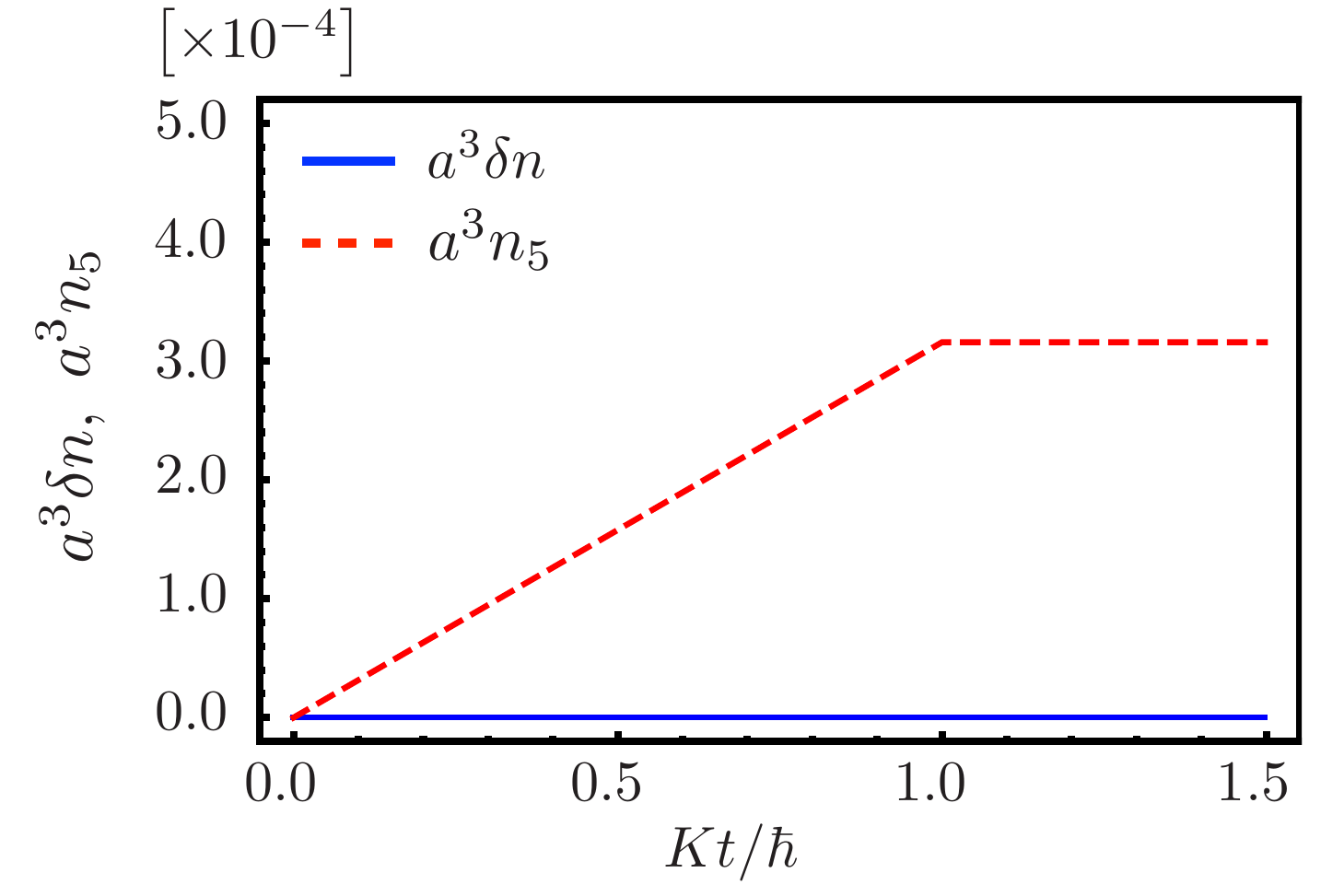}
\caption{\label{fig4:density}
Time evolution of the number-density fluctuation $\delta n$ and that of the chiral density $n_5$. 
}
\end{figure}

To look at the spectral flow more closely, we have calculated the number density $n$ and the chiral density $n_5$.
We define the chiral density by dividing the first Brillouin zone into the eight regions so that the eight Weyl nodes are located at their centers. 
Then $n$ and $n_5$ are obtained from the momentum integration of $\sqrt{\omega}n_{\bm p}$ in each region, $n_{\bm{p}_w}$, as $n=\sum_{\bm{p}_w} n_{\bm{p}_w}$, and $n_5=\sum_{\bm{p}_w}\kappa_{\bm{p}_w} n_{\bm{p}_w}$. 
We show the deviation of the number density from its initial value $\delta n(t)=n(t)-n(0)$ in Fig.~\ref{fig4:density}.
Our simulation satisfies the particle-number conservation.
We also show the chiral density in Fig.~\ref{fig4:density}. We see that the chiral density increases as the spectral flows grows, which induces the chiral magnetic current as shown below.

\begin{figure}[t]
\centering
\includegraphics[scale=0.55]{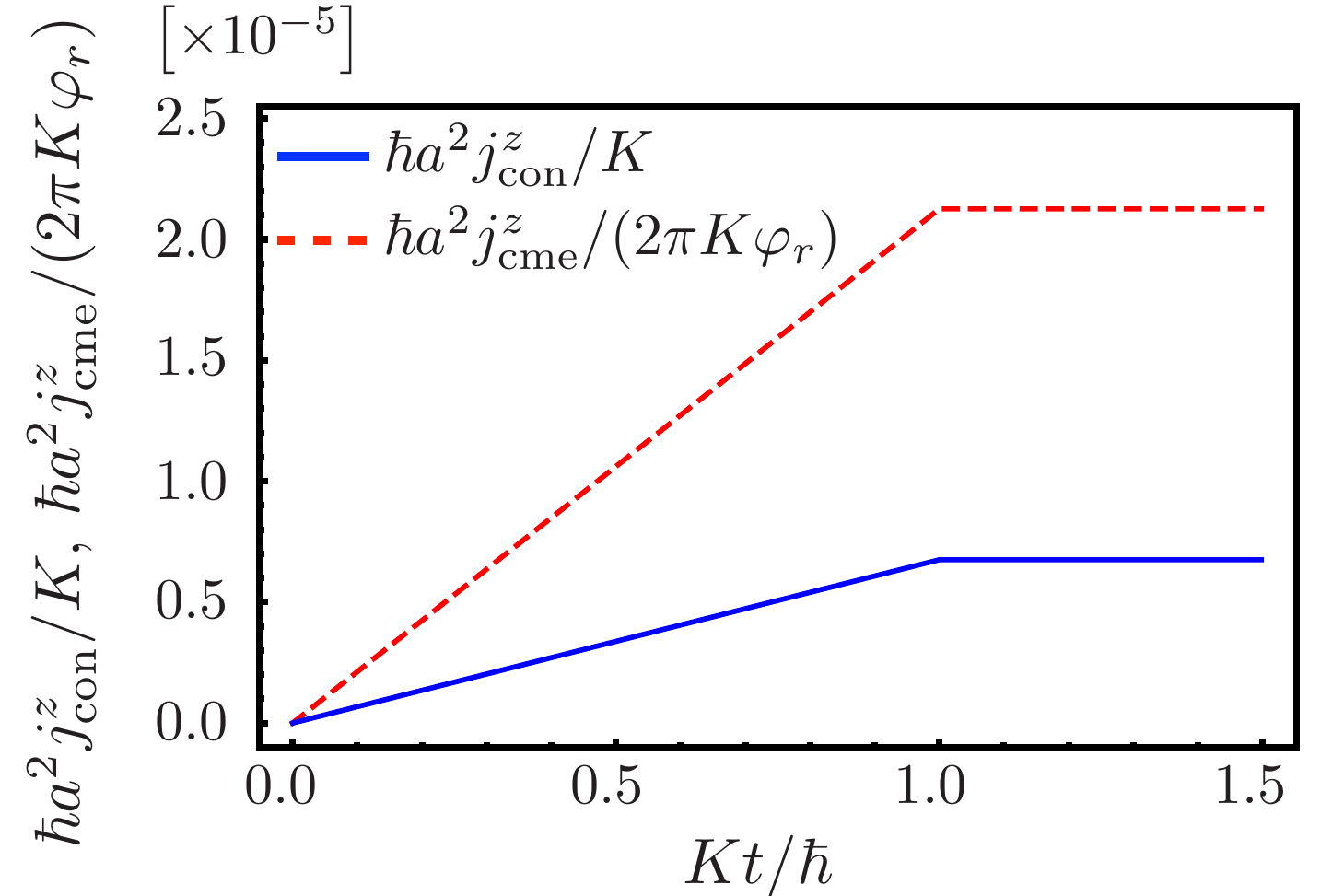}
\caption{\label{fig5:cme}
Time evolution of the conventional current density $\bm{j}_{\rm con}$ and that of the chiral magnetic current density $\bm{j}_{\rm cme}$. 
The anomalous Hall current density $\bm{j}_{\rm Hall}$ and other components not shown in the figure vanish at all times.
}
\end{figure}

We show the chiral magnetic current in Fig.~\ref{fig5:cme}.
The current density reads $\bm{j}=\int d^3p/(2\pi)^3\sqrt{\omega}\dot{\bm x}n_{\bm p}=
\int d^3p/(2\pi)^3(\bm v+\frac{e}{\hbar}\bm E\times\bm \Omega+\left(\bm v\cdot\bm \Omega\right)\frac{e}{\hbar c}\bm B)n_{\bm p}$~\cite{Stephanov:2012ki,PhysRevB.88.104412}.
The first term on the right-hand side is the conventional current $\bm{j}_{\rm con}$. The second is the the anomalous Hall current $\bm{j}_{\rm Hall}$, which vanishes since $n_{\bm p}$ is isotropic in the $p_x$-$p_y$ plane around the Weyl nodes as seen in Fig.~\ref{fig3:ring}.
The last is the chiral magnetic current $\bm{j}_{\rm cme}$.
The first and last terms can be expressed by using $\nabla_{\bm p}n_{\bm p}$ to make the contributions from the Fermi surface manifest~\cite{Son:2012wh}.
We have confirmed that both expressions give the same result. 
We find that as the spectral flow grows, the chiral magnetic current increases, and once the spectral flow ceases, so does the chiral magnetic current, 
which is consistent with Fig.~\ref{fig4:density}.

{\it Concluding remarks.} 
We have analyzed the spectral flow in a Weyl metal constructed from a three-dimensional optical lattice.
By numerically solving the Boltzmann equation, which involves the Berry curvature in the presence of synthetic electric and magnetic fields on the basis of the CIP scheme,
we have successfully simulated the spectral flow, and shown that the particle is excited from the left-handed Weyl nodes to the right-handed ones through the triangle anomaly only if $\bm{E}\cdot\bm{B}$ is nonzero. 
As a consequence, the Fermi surface around the right-handed (left-handed) Weyl node enlarges (shrinks) and the chiral chemical potential is dynamically generated.
The difference between the Fermi surfaces can be experimentally observed as the double ring pattern by the time-of-flight imaging with adiabatic ramping.
Also we have analyzed the time evolution of the chiral magnetic current.

There are several future applications.
We can analyze the effect of dissipation by directly applying our simulation on the basis of the relaxation time approximation~\cite{PhysRevB.88.104412}.
By considering dissipation, we can simulate a nonequilibrium steady state with a nonzero chiral chemical potential, 
where the excitation via the triangle anomaly is balanced by  dissipation. 

Another possible direction is a simulation in real space coordinates.
Since there are nonzero currents, particles move in real space. 
To fully understand the nonequilibrium physics, we need to solve the dynamics in real space.
However a simulation in full six-dimensional coordinates requires a huge computational cost.
Our approach is also applicable to other anomalous transports induced via the triangle anomaly.
It is also of interest to solve the Boltzmann equation under rotation~\cite{PhysRevB.89.035142} or dislocation~\cite{PhysRevLett.116.166601}.

Our analysis can be applied to relativistic systems as well as condensed matter systems. 
We can study the dynamical evolution of the chiral magnetic/vortical current and estimate their effects on heavy ion collision experiments~\cite{Kharzeev:2015znc}, measurements of neutron stars (magnetars)~\cite{Ohnishi:2014uea} and neutrino physics in the early universe~\cite{Yamamoto:2015gzz}.

\begin{acknowledgements}
T.~H. thanks Y.~Hidaka, Y.~Tachibana, Y.~Tanizaki, N.~Tsuji, S.~Uchino, and N.~Yamamoto for stimulating discussions.
T.~H. is supported by Grants-in-Aid for the fellowship of Japan Society for the Promotion
of Science (JSPS) (No: JP16J02240).
This work was supported by
KAKENHI Grant No. 26287088 from the Japan Society for the Promotion of Science, 
a Grant-in-Aid for Scientific Research on Innovative Areas ``Topological Materials Science’’ (KAKENHI Grant No. 15H05855), 
the Photon Frontier Network Program from MEXT of Japan,
and the Mitsubishi Foundation.

\end{acknowledgements}


\bibliography{./weyl}

\end{document}